# Risk management in the Artificial Intelligence Act

*Jonas Schuett\**

The proposed EU AI Act is the first comprehensive attempt to regulate AI in a major jurisdiction. This article analyses Article 9, the key risk management provision in the AI Act. It gives an overview of the regulatory concept behind Article 9, determines its purpose and scope of application, offers a comprehensive interpretation of the specific risk management requirements, and outlines ways in which the requirements can be enforced. This article is written with the aim of helping providers of high-risk systems comply with the requirements set out in Article 9. In addition, it can inform revisions of the current draft of the AI Act and efforts to develop harmonised standards on AI risk management.

## I. Introduction

In April 2021, the European Commission (EC) published a proposal for an Artificial Intelligence Act (AI Act).[1] As the first comprehensive attempt to regulate[2] artificial intelligence (AI)[3] in a major jurisdiction, the AI Act will

---

\* Research Fellow, Centre for the Governance of AI, Oxford, UK; Research Affiliate, Legal Priorities Project, Cambridge, MA, USA; PhD Candidate, Faculty of Law, Goethe University Frankfurt, Germany; jonas.schuett@governance.ai.

[1] EC, 'Proposal for a Regulation of the European Parliament and of the Council Laying Down Harmonised Rules on Artificial Intelligence (Artificial Intelligence Act) and Amending Certain Union Legislative Acts' COM (2021) 206 final <https://perma.cc/4YXM-38U9>. Unless otherwise specified, my analysis refers to the text of the original proposal and not to the amendments advanced so far in the legislative process.

[2] The term "regulation" can be defined as "sustained and focused attempts to change the behaviour of others in order to address a collective problem or attain an identified end or ends, usually but not always through a combination of rules or norms and some means for their implementation and enforcement, which can be legal or non-legal" (Julia Black and Andrew Murray, 'Regulating AI and Machine Learning: Setting the Regulatory Agenda' [2019] 10 European Journal of Law and Technology <https://perma.cc/A456-QPHH>). For a collection of definitions, see Christel Koop and Martin Lodge, 'What Is Regulation? An Interdisciplinary Concept Analysis' (2017) 11 Regulation & Governance 95 <https://doi.org/10.1111/rego.12094>.

[3] There is no generally accepted definition of the term "AI". Since its first usage by John McCarthy and others, 'A Proposal for the Dartmouth Summer Research Project on Artificial Intelligence' (1955) <https://perma.cc/PEK4-MKHF>, a vast spectrum of definitions has

inevitably serve as a benchmark for other countries like the United States (US) and the United Kingdom (UK). Due to the so-called "Brussels Effect",[4] it might even have *de facto* effects in other countries,[5] similar to the General Data Protection Regulation (GDPR).[6] It will undoubtedly shape the foreseeable future of AI regulation in the European Union (EU) and worldwide.

Within the AI Act, the requirements on risk management[7] are particularly important. AI can cause or exacerbate a wide range of risks, including

---

emerged. For a collection of definitions, see Shane Legg and Marcus Hutter, 'A Collection of Definitions of Intelligence' (arXiv, 2007) <https://arxiv.org/abs/0706.3639>; Sofia Samoili and others, 'AI Watch: Defining Artificial Intelligence: Towards an Operational Definition and Taxonomy of Artificial Intelligence' (2020) <https://doi.org/10.2760/382730>. Categorizations of different AI definitions have been proposed by Stuart J Russell and Peter Norvig, *Artificial Intelligence: A Modern Approach* (4th edn, Pearson 2021); Pei Wang, 'On Defining Artificial Intelligence' (2019) 10 Journal of Artificial General Intelligence 1 <https://doi.org/10.2478/jagi-2019-0002>; Sankalp Bhatnagar and others, 'Mapping Intelligence: Requirements and Possibilities' in Vincent C Müller (ed), *Philosophy and Theory of Artificial Intelligence 2017* (Springer International Publishing 2018) <https://doi.org/10.1007/978-3-319-96448-5_13>. For a discussion of the term in a regulatory context, see Jonas Schuett, 'Defining the Scope of AI Regulations' 15 Law, Innovation and Technology (forthcoming) <https://arxiv.org/abs/1909.01095>. Art. 3, point 1 defines an "AI system" as "software that is developed with one or more of the techniques and approaches listed in Annex I and can, for a given set of human-defined objectives, generate outputs such as content, predictions, recommendations, or decisions influencing the environments they interact with".

[4] The term "Brussels Effect" has been coined by Anu Bradford, 'The Brussels Effect' (2012) 107 Northwestern University Law Review 1 <https://perma.cc/SK85-T2QM>; see also Anu Bradford, *The Brussels Effect: How the European Union Rules the World* (OUP 2020).

[5] See Charlotte Siegmann and Markus Anderljung, 'The Brussels Effect and Artificial Intelligence: How EU Regulation Will Impact the Global AI Market' (Centre for the Governance of AI 2022) <https://perma.cc/VS8H-P96U>; Alex Engler, 'The EU AI Act Will Have Global Impact, but a Limited Brussels Effect' (Brookings Institution 2022) <https://perma.cc/YYH4-83QU>.

[6] Regulation (EU) 2016/679 of the European Parliament and of the Council of 27 April 2016 on the protection of natural persons with regard to the processing of personal data and on the free movement of such data, and repealing Directive 95/46/EC (General Data Protection Regulation) [2016] OJ L119/1.

[7] The term "risk management" can be defined as the "coordinated activities to direct and control an organisation with regard to risk" (Clause 3.2 of 'ISO 31000:2018 Risk Management — Guidelines' <https://www.iso.org/standard/65694.html> accessed 2 November 2022).

accident,[8] misuse,[9] and structural risks.[10] Organisations that develop and deploy AI systems need to manage these risks for economic, legal, and ethical reasons. Being able to reliably identify, accurately assess, and adequately respond to risks from AI is particularly important in high-stakes situations (e.g. if AI systems are used in critical infrastructure[11]). This will become even more important as AI systems become more capable and more general in the future.[12]

In recent years, attention on AI risk management has increased steadily amongst practitioners. As of 2022, several standard-setting bodies are developing voluntary AI risk management frameworks; the most notable ones are the NIST AI Risk Management Framework[13] and ISO/IEC FDIS 23894.[14]

---

[8] For more information on accident risks, see Dario Amodei and others, 'Concrete Problems in AI Safety' (arXiv, 2016) <https://arxiv.org/abs/1606.06565>; Zachary Arnold and Helen Toner, 'AI Accidents: An Emerging Threat' (Center for Security and Emerging Technology 2021) <https://perma.cc/V2AY-PFY5>.

[9] For more information on misuse risks (also referred to as "malicious use"), see Miles Brundage and others, 'The Malicious Use of Artificial Intelligence: Forecasting, Prevention, and Mitigation' (arXiv, 2018) <https://arxiv.org/abs/1802.07228>.

[10] For more information on structural risks, see Remco Zwetsloot and Allan Dafoe, 'Thinking About Risks From AI: Accidents, Misuse and Structure' (*Lawfare*, 11 February 2019) <https://perma.cc/H3CQ-SEQ9>.

[11] E.g. in early 2022, DeepMind announced a breakthrough in using AI in nuclear fusion reactors (Jonas Degrave and others, 'Magnetic Control of Tokamak Plasmas through Deep Reinforcement Learning' [2022] 602 Nature 414 <https://doi.org/10.1038/s41586-021-04301-9>).

[12] Forecasting AI progress is an inherently difficult endeavour which involves substantial methodological difficulties. One approach is to survey the views of leading AI researchers (see e.g. Katja Grace and others, 'Viewpoint: When Will AI Exceed Human Performance? Evidence from AI Experts' [2018] 62 Journal of Artificial Intelligence Research 729 <https://doi.org/10.1613/jair.1.11222>; Baobao Zhang and others, 'Forecasting AI Progress: Evidence from a Survey of Machine Learning Researchers' [arXiv, 2022] <https://arxiv.org/abs/2206.04132>; Zach Stein-Perlman, Benjamin Weinstein-Raun and Katja Grace, '2022 Expert Survey on Progress in AI' [*AI Impacts*, 3 August 2022] <https://perma.cc/CE2L-PRAA>). Another approach is to extrapolate current AI trends, e.g. that using more data (Michael I Jordan and Tom M Mitchell, 'Machine Learning: Trends, Perspectives, and Prospects' [2015] 349 Science <https://doi.org/10.1126/science.aaa8415>) and more compute (Jaime Sevilla and others, 'Compute Trends Across Three Eras of Machine Learning' [arXiv, 2022] <https://arxiv.org/abs/2202.05924>) to train bigger models (Pablo Villalobos and others, 'Machine Learning Model Sizes and the Parameter Gap' [arXiv, 2022] <https://arxiv.org/abs/2207.02852>) leads to improved capabilities (Jared Kaplan and others, 'Scaling Laws for Neural Language Models' [arXiv, 2020] <https://arxiv.org/abs/2001.08361>).

[13] NIST, 'AI Risk Management Framework: Second Draft' <https://perma.cc/6EJ9-UZ9A>.

[14] 'ISO/IEC FDIS 23894 Information Technology — Artificial Intelligence — Guidance on Risk Management' <https://www.iso.org/standard/77304.html> accessed 2 November 2022.

Existing enterprise risk management (ERM) frameworks like COSO ERM 2017[15] have also been applied to an AI context.[16] Many consulting firms have published reports on AI risk management.[17] However, there is only limited academic literature on the topic.[18] In particular, there does not seem to be any literature analysing the risk management provision in the AI Act.[19]

This article conducts a doctrinal analysis[20] of Article 9 using the four methods of statutory interpretation: literal, systematic, teleological, and historical interpretation.[21] But since there is not yet a final text, I have to rely on drafts and proposals,[22] namely the original draft by the EC,[23] as well as the proposed changes by the French[24] and Czech Presidency of the Council[25] and the European Parliament (EP).[26]

---

[15] COSO, 'Enterprise Risk Management—Integrating with Strategy and Performance' (2017) <https://perma.cc/G6JD-BBWB>.

[16] E.g. Keri Calagna, Brian Cassidy and Amy Park, 'Realizing the Full Potential of Artificial Intelligence - Applying the COSO ERM Framework and Principles to Help Implement and Scale AI' (2021) <https://perma.cc/SD7Z-9XPU>.

[17] E.g. Benjamin Cheatham, Kia Javanmardian and Hamid Samandari, 'Confronting the Risks of Artificial Intelligence' (McKinsey 2019) <https://perma.cc/T2CX-HYZF>; PwC, 'Model Risk Management of AI and Machine Learning Systems' (2020) <https://perma.cc/RBC2-BHZN>; Gabriella Ezeani and others, 'A Survey of Artificial Intelligence Risk Assessment Methodologies - The Global State of Play and Leading Practices Identified' (EY 2022) <https://perma.cc/WRD7-5JPV>.

[18] E.g. Gergő Barta and Gergely Görcsi, 'Risk Management Considerations for Artificial Intelligence Business Applications' (2021) 21 International Journal of Economics and Business Research 87 <https://dx.doi.org/10.1504/IJEBR.2021.10031075>; Robin Nunn, 'Discrimination in the Age of Algorithms' in Woodrow Barfield (ed), *The Cambridge Handbook of the Law of Algorithms* (CUP 2020) 195 <https://doi.org/10.1017/9781108680844.010>; Alette Tammenga, 'The Application of Artificial Intelligence in Banks in the Context of the Three Lines of Defence Model' (2020) 94 Maandblad Voor Accountancy en Bedrijfseconomie 219 <https://doi.org/10.5117/mab.94.47158>. See also related work by Luca Enriques and Dirk A Zetzsche, 'The Risky Business of Regulating Risk Management in Listed Companies' (2013) 103 European Company and Financial Law Review 271 <http://dx.doi.org/10.2139/ssrn.2344314>.

[19] The only exceptions seem to be Tobias Mahler, 'Between Risk Management and Proportionality: The Risk-Based Approach in the EU's Artificial Intelligence Act Proposal' [2022] Nordic Yearbook of Law and Informatics 247 <https://doi.org/10.53292/208f5901.38a67238> who focuses on the general approach, not the provision itself, and a short blog post by Mark Cankett and Barry Liddy, 'Risk Management in the New Era of AI Regulation - Considerations around Risk Management Frameworks in Line with the Proposed EU AI Act' (*Deloitte*, 12 July 2022) <https://perma.cc/2W95-J67Z>.

[20] For more information on doctrinal legal research, see Terry Hutchinson and Nigel Duncan, 'Defining and Describing What We Do: Doctrinal Legal Research' (2012) 17 Deakin Law Review 83 <https://doi.org/10.21153/dlr2012vol17no1art70>.

[21] For more information on the interpretation of EU law, see Koen Lenaerts and José A Gutiérrez-Fonz, 'To Say What the Law of the EU Is: Methods of Interpretation and the European Court of Justice' (European University Institute 2013) <https://perma.cc/2XZN-RAH8>; see also Reinhold Zippelius and Thomas Würtenberger, *Juristische Methodenlehre* (12th edn, CH Beck 2021).

Since my analysis relies on drafts and proposals, it is possible that future changes will make my analysis obsolete. However, there are three main reasons why I am willing to take that risk. First, I do not expect the provision to change significantly. The requirements are fairly vague and do not seem to be that controversial. In particular, I am not aware of significant public debates about Article 9 (although the French[27] and Czech Presidency of the Council[28] as well as the EP[29] have suggested changes). Second, even if the provision is changed, it seems unlikely that the whole analysis would be affected. Most parts would probably remain relevant. Section III, which determines the purpose of the provision, seems particularly robust to future changes. Third, in some cases, changes might even be desirable. In Sections VII, I suggest several amendments myself. In short, I would rather publish my analysis too early than too late.

The article proceeds as follows. Section II gives an overview of the regulatory concept behind Article 9. Section III determines its purpose and Section IV its scope of application. Section V contains a comprehensive interpretation of the specific risk management requirements, while Section VI outlines ways in which they can be enforced. Section VII concludes with recommendations for the further legislative process.

---

[22] For an up-to-date list with relevant documents, see Kai Zenner, 'Documents and Timelines: The Artificial Intelligence Act (Part 3)' (*Digitizing Europe*, 12 October 2022) <https://www.kaizenner.eu/post/aiact-part3> accessed 1 November 2022.

[23] COM (2021) 206 final, supra, note 1.

[24] French Presidency of the Council, 'Presidency Compromise Text - Consolidated Version' (2022) <https://perma.cc/VF59-CTJK>.

[25] Czech Presidency of the Council, 'Final Presidency Compromise Text' (2022) <https://perma.cc/9YY7-JQ3D>.

[26] Within the EP, the process is led by two committees that have proposed amendments (Committee on the Internal Market and Consumer Protection [IMCO] and Committee on Civil Liberties, Justice and Home Affairs [LIBE], 'Draft Report' [2022] <https://perma.cc/AC4G-T6SN>). The opinions of five other committees have to be taken into account (Committee on Legal Affairs [JURI], 'Opinion' [2022] <https://perma.cc/K4P5-KJ5M>; Committee on Industry, Research and Energy [ITRE], 'Opinion' [2022] <https://perma.cc/G6P3-SPB6>; Committee on Culture and Education [CULT], 'Opinion' [2022] <https://perma.cc/8XME-MUVA>; Committee on the Environment, Public Health and Food Safety [ENVI], 'Opinion' [2022] <https://perma.cc/BZD9-S3ZM>; Committee on Transport and Tourism [TRAN], 'Opinion' [2022] <https://perma.cc/V83P-WWRJ>).

[27] French Presidency of the Council, supra, note 24.

[28] Czech Presidency of the Council, supra, note 25.

[29] IMCO and LIBE, 'All Amendments' (2022) <https://perma.cc/W7ZL-AJYJ>.

## II. Regulatory concept

In this section, I give an overview of the regulatory concept behind Article 9. I analyse its role in the AI Act, its internal structure, and the role of standards.

The AI Act famously takes a risk-based approach.[30] It prohibits AI systems with unacceptable risks[31] and imposes specific requirements on high-risk AI systems,[32] while leaving AI systems that pose low or minimal risks largely unencumbered.[33] To reduce the risks from high-risk AI systems, providers of such systems must comply with the requirements set out in Chapter 2,[34] but the AI Act assumes that this will not be enough to reduce all risks to an acceptable level: even if providers of high-risk AI systems comply with the requirements, some risks will remain. The role of Article 9 is to make sure that providers identify those risks and take additional measures to reduce them to an acceptable level.[35] In this sense, Article 9 serves an important backup function.

The norm is structured as follows. Paragraph 1 contains the central requirement, according to which providers of high-risk AI systems must implement a risk management system, while paragraphs 2 to 7 specify the details of that system. The key element of the risk management system, the risk management process, is described in paragraph 2. The remainder of Article 9 contains special rules about risk management measures (paragraphs 3 and 4), testing procedures (paragraphs 5 to 7), and children and credit institutions (paragraphs 8 and 9).

In the regulatory concept of the AI Act, standards play a key role.[36] By complying with harmonised standards,[37] regulatees can demonstrate compliance with the requirements set out in the AI Act.[38] This effect is called "presumption of conformity".[39] In areas where no harmonised standards exist or where they are insufficient, the EC can also develop common specifications.[40] Harmonised standards and common specifications are explicitly mentioned in Article 9(3), sentence 2. It is worth noting that the French Presidency of the Council has suggested deleting the reference to harmonised standards and common specifications,[41] and the Czech Presidency has adopted that suggestion.[42] However, this would not undermine the importance of harmonised standards and common specifications. They would continue to provide guidance and presume conformity. Harmonised standards and common specifications on AI risk management do not yet exist. The recognised European Standards Organisations[43] have jointly been tasked with creating technical standards for the AI Act, including risk management systems,[44] but that process may take several years. In the meantime, regulatees could use international standards like the NIST AI Risk Management Framework[45] or ISO/IEC DIS 23894.[46] Although this will not presume conformity, these standards can still serve as a rough guideline. In particular, I expect them to be similar to the ones that will be created by the European Standards Organizations, mainly because standard-setting efforts usually strive for some level of compatibility,[47] but of course, there is no guarantee for this. With this regulatory concept in mind, let us now take a closer look at the purpose of Article 9.

---

[30] See Recital 14. Risk-based regulation is a regulatory approach that tries to achieve policy objectives by targeting activities that pose the highest risk, while lowering the burdens for low-risk activities (see Julia Black, 'Risk-Based Regulation: Choices, Practices and Lessons Being Learnt' in OECD [ed], *Risk and Regulatory Policy: Improving the Governance of Risk* [2010] 187 <https://doi.org/10.1787/9789264082939-en>; Robert Baldwin and Julia Black, 'Driving Priorities in Risk-Based Regulation: What's the Problem?' [2016] 43 Journal of Law and Society 565, 565 <https://doi.org/10.1111/jols.12003>). For more information on the risk-based approach in the AI Act, see Mahler, supra, note 19.

[31] Art. 5.

[32] Art. 9–15.

[33] See Art. 52.

[34] Art. 8 and 16(a). Chapter 2 contains requirements on risk management (Art. 9), data and data governance (Art. 10), technical documentation (Art. 11), record-keeping (Art. 12), transparency and the provision of information to users (Art. 13), human oversight (Art. 14), and accuracy, robustness and cybersecurity (Art. 15).

[35] See Sections V.1 and V.2.

[36] See Recital 61, sentence 1. For more information on the role of standards in the AI Act, see Marc McFadden and others, 'Harmonising Artificial Intelligence - The Role of Standards in the EU AI Regulation' (Oxford Information Labs 2021) <https://perma.cc/X3AZ-5H7C>.

[37] The term "harmonised standard" is defined in Art. 3, point 27 in conjunction with Art. 2(1), point (c) of Regulation (EU) No 1025/2012.

## III. Purpose

In this section, I determine the purpose of Article 9. This is an important step, because the purpose has significant influence on the extent to which different interpretations of the provision are permissible.

---

[38] See Art. 40.
[39] See Art. 65(6), sentence 2, point (b).
[40] See Art. 41. The term "common specification" is defined in Art. 3, point 28.
[41] French Presidency of the Council, supra, note 24.
[42] Czech Presidency of the Council, supra, note 25.
[43] The European Committee for Standardization (CEN), the European Committee for Electrotechnical Standardization (CENELEC), and the European Telecommunications Standards Institute (ETSI).
[44] Luca Bertuzzi, 'AI Standards Set for Joint Drafting among European Standardisation Bodies' (*Euractiv*, 30 May 2022) <https://perma.cc/3VB6-CHRX>.
[45] NIST, supra, note 13.
[46] 'ISO/IEC FDIS 23894 Information Technology — Artificial Intelligence — Guidance on Risk Management', supra, note 14.
[47] See the statement by the US and EU, 'Joint Statement of the Trade and Technology Council' (2022) 9 <https://perma.cc/2F57-23J9>. See also McFadden and others, supra, note 36, 14.

Pursuant to Recital 1, sentence 1, the purpose of the AI Act is "to improve the functioning of the internal market by laying down a uniform legal framework […] in conformity with Union values." More precisely, the AI Act intends to improve the functioning of the internal market through preventing fragmentation and providing legal certainty.[48] The legal basis for this is Article 114 of the Treaty on the Functioning of the European Union (TFEU).[49]

At the same time, the AI Act is intended to ensure a "high level of protection of public interests".[50] Relevant public interests include "health and safety and the protection of fundamental rights, as recognised and protected by Union law".[51] Note that the French Presidency of the Council has suggested adding a reference to "health, safety and fundamental rights" in Article 9(2), sentence 2, point (a),[52] which the Czech Presidency has adopted.[53] Protecting these public interests is part of the EU's objective of becoming a leader in "secure, trustworthy and ethical artificial intelligence".[54]

It is unclear if Article 9 is also intended to protect individuals. This would be important because, if it does, it would be easier for the protected individuals to assert tort claims in certain member states.[55] Recital 42 provides an argument in favour. It states that the requirements for high-risk AI systems are intended to mitigate the risks to users[56] and affected persons.[57] However, one could also hold the view that the risk management system is primarily an organisational requirement that only indirectly affects individuals.[58] Since this question is beyond the scope of this article, I will leave it open.

---

[48] See Recital 2, sentences 3 and 4; see also Recital 1, sentence 2.

[49] Recital 2, sentence 4, but note the exception for biometric identification in Recital 2, sentence 5.

[50] See Recital 2, sentence 4.

[51] Recital 5, sentence 1 and Recital 1, sentence 2; see also the Charter of Fundamental Rights of the European Union.

[52] French Presidency of the Council, supra, note 24.

[53] The Czech Presidency of the Council, supra, note 25.

[54] Recital 5, sentence 3.

[55] E.g. Section 823(2) of the German Civil Code.

[56] The term "user" is defined in Art. 3, point 4.

[57] The AI Act does not define the term "affected person". "Person" could refer to any natural or legal person, similar to the definition of "user" in Art. 3, point 4. Other EU regulations that use the term also define it with reference to both natural and legal persons (see e.g. Art. 2, point 10 of Regulation [EU] 2018/1805). However, the definition could also be limited to natural persons, as implied by a statement in the proposal, according to which Title III, including Art. 9, is concerned with "high risk to the health and safety or fundamental rights of natural persons" (COM [2021] 206 final, supra, note 1, 13). Since this question is beyond the scope of this article, I will leave it open. A person is "affected", if they are subject to the adverse effects of an AI system. Note that the AI Act pays special attention to adverse effects on health, safety and fundamental rights (see Recital 1, sentence 2).

[58] This seems to be assumed by Art. 4(2) of EC, 'Proposal for a Regulation of the European Parliament and of the Council on Adapting Non-Contractual Civil Liability Rules to

Understanding the purpose of Article 9 helps interpreting the specific risk management requirements. But before we can turn to that, we must first determine who needs to comply with these requirements.

## IV. Scope of application

In this section, I determine the scope of Article 9. This includes the material scope (what is regulated), the personal scope (who is regulated), the regional scope (where the regulation applies), and the temporal scope (when the regulation applies).[59]

Article 9 only applies to "high-risk AI systems". This can be seen by the formulation in paragraph 1 ("in relation to high-risk AI systems") and the location of Article 9 in Chapter 2 ("Requirements for high-risk AI systems"). The term "AI system" is defined in Article 3, point 1,[60] while Article 6 and Annex III specify which AI systems qualify as high-risk. This includes, for example, AI systems that screen or filter applications as well as risk assessment tools used by law enforcement authorities. Note that both the French[61] and Czech Presidency[62] as well as the EP[63] have suggested changes to the AI definition.

The risk management system does not need to cover AI systems that pose unacceptable risks; these systems are prohibited.[64] But what about AI systems that pose low or minimal risks? Although there is no legal requirement to include such systems, I would argue that, in many cases, it makes sense to do so on a voluntary basis. There are at least two reasons for this. First, if organisations want to manage risks holistically,[65] they should not exclude certain risk categories from the beginning. The risk classification in the AI Act does not guarantee that systems below the high-risk threshold do not pose any other

---

Artificial Intelligence (AI Liability Directive)' COM (2022) 496 final <https://perma.cc/54M5-V8YB>, which facilitates tort claims for individuals in case of violations of many provisions of Title III, Chapter 2 of the AI Act, but not Art. 9.

[59] For more information on defining the scope of AI regulations, see Schuett, supra, note 3.

[60] The term "AI system" is defined as "software that is developed with one or more of the techniques and approaches listed in Annex I and can, for a given set of human-defined objectives, generate outputs such as content, predictions, recommendations, or decisions influencing the environments they interact with".

[61] French Presidency of the Council, supra, note 24.

[62] Czech Presidency of the Council, supra, note 25.

[63] IMCO and LIBE, supra, note 29.

[64] See Art. 5.

[65] This is the key characteristic of ERM, see e.g. Philip Bromiley and others, 'Enterprise Risk Management: Review, Critique, and Research Directions' (2015) 48 Long Range Planning 265 <https://doi.org/10.1016/j.lrp.2014.07.005>.

risks that are relevant to the organisation, such as litigation and reputation risks. It therefore seems preferable to initially include all risks. After risks have been identified and assessed, organisations can still choose not to respond. Second, most of the costs for implementing the risk management system will likely be fixed costs, which means that including low and minimal-risk AI systems would only marginally increase the operating costs.

In addition, both the French[66] and the Czech Presidency of the Council[67] have suggested extending Article 9 to "general purpose AI systems".[68] Meanwhile, the amendments under consideration by the EP range from extending Article 9 to general purpose AI systems to completely excluding them from the scope of the AI Act.[69] Overall, the best approach to regulating general purpose AI systems is still highly disputed and beyond the scope of this article.[70]

Since Article 9 is formulated in the passive voice ("a risk management system shall be established"), it does not specify who needs to comply with the requirements. However, Article 16, point (a) provides clarity: Article 9 only applies to "providers of high-risk AI systems". The term "provider" is defined in Article 3, point 2.[71] Note that Article 2(4) excludes certain public authorities and international organisations from the personal scope.

Article 9 has the same regional scope as the rest of the AI Act. According to Article 2(1), the AI Act applies to providers who place on the market[72] or put into service[73] AI systems in the EU, or where the output produced by AI systems is used in the EU. It does not matter if the provider of such systems is established within the EU or in a third country. The regional scope of the AI

---

[66] French Presidency of the Council, supra, note 24.

[67] Czech Presidency of the Council, supra, note 25.

[68] In these documents, the term "general purpose AI system" is defined as "an AI system that - irrespective of how it is placed on the market or put into service, including as open source software - is intended by the provider to perform generally applicable functions such as image and speech recognition, audio and video generation, pattern detection, question answering, translation and others; a general purpose AI system may be used in a plurality of contexts and be integrated in a plurality of other AI systems".

[69] IMCO and LIBE, supra, note 29.

[70] See e.g. Alex Engler, 'To Regulate General Purpose AI, Make the Model Move' (*Tech Policy Press*, 10 November 2022) <https://perma.cc/6J8X-C7GT>. General purpose AI systems may warrant special risk management implementation. Thus, according to the proposal by the Czech Presidency of the Council, supra, note 25, implementing acts by the Commission "shall specify and adapt the application of the requirements established in Title III, Chapter 2 to general purpose AI systems in the light of their characteristics, technical feasibility, specificities of the AI value chain and of market and technological developments."

[71] The term "provider" is defined as "a natural or legal person, public authority, agency or other body that develops an AI system or that has an AI system developed with a view to placing it on the market or putting it into service under its own name or trademark, whether for payment or free of charge".

[72] The term "placing on the market" is defined in Art. 3, point 9.

[73] The term "putting into service" is defined in Art. 3, point 11.

Act is relatively broad. The EC justifies this with the "digital nature" of AI systems.[74]

Providers of high-risk AI systems must have implemented a risk management system 24 months after the AI Act enters into force.[75] (The French Presidency of the Council has proposed to extend this period to 36 months.[76] The Czech Presidency has adopted that proposal.[77]) The AI Act will enter into force 20 days after its publication in the Official Journal of the European Union. It is unclear when this will be the case. The EC is currently waiting for the Council and the EP to finalise their positions. The Council's position will likely be put forward by the Czech Presidency in late 2022 or the Swedish Presidency in early 2023. Similarly, the EP will not vote on their position before the end of 2022 or early 2023. Once the Council and EP have finalised their positions, they will enter interinstitutional negotiations assisted by the EC, the so-called "trilogue". Against this background, it seems unlikely that the final regulation will enter into force before early 2023. Providers of high-risk AI systems therefore have time until early 2025 (or 2026 according to the proposal by the French and the Czech Presidency of the Council[78]) to comply with the requirements set out in Article 9. But what exactly do these requirements entail?

## V. Requirements

In this section, I offer a comprehensive interpretation of the specific risk management requirements set out in Article 9.

### 1. Risk management system, Article 9(1)

Pursuant to paragraph 1, "a risk management system shall be established, implemented, documented and maintained in relation to high-risk AI systems." This is the central requirement of Article 9.

The AI Act does not define the term "risk management system",[79] but the formulation in paragraph 8 suggests that it means all measures described in paragraphs 1 to 7, namely the risk management process (paragraphs 2 to 4) and testing procedures (paragraphs 5 to 7). Analogous to the description of the quality management system in Article 17(1), one could hold the view that a "system" consists of policies, procedures, and instructions.

---

[74] See Recital 11.
[75] See Art. 85(2).
[76] French Presidency of the Council, supra, note 24.
[77] Czech Presidency of the Council, supra, note 25.
[78] French Presidency of the Council, supra, note 24.
[79] The term "risk management" can be defined as "coordinated activities to direct and control an organisation with regard to risk" (Clause 3.2 of 'ISO 31000:2018 Risk Management — Guidelines', supra, note 7).

The risk management system needs to be "established, implemented, documented and maintained". Since none of these terms are defined in the AI Act, I suggest the following definitions. A risk management system is "established" if risk management policies, procedures and instructions are created[80] and approved by the responsible decision-makers.[81] It is "implemented" if it is put into practice, i.e. the employees concerned understand what is expected of them and act accordingly.[82] It is "documented" if the system is described in a systematic and orderly manner in the form of written policies, procedures and instructions,[83] and can be demonstrated upon request of a national competent authority.[84] It is "maintained" if it is reviewed and, if necessary, updated on a regular basis.[85]

The risk management system must be established "in relation to high-risk AI systems".[86] This means that the system only needs to cover risks from high-risk AI systems. Inversely, it does not have to address risks from AI systems that pose low or minimal risks. However, as I have argued in Section IV, it might make sense for an organisation to do so on a voluntary basis.

## 2. Risk management process, Article 9(2)

The first component of the risk management system is the risk management process. This process specifies how providers of high-risk AI systems must identify, assess, and respond to risks. Paragraph 2 defines the main steps of this process, while paragraphs 3 and 4 contain further details about specific risk management measures.[87] Note that most terms are not defined in the AI Act, but since the risk management process in the AI Act seems to be inspired by ISO/IEC Guide 51,[88] I use or adapt many of their definitions.

---

[80] In practice, I expect many providers of high-risk AI systems to seek advice from consulting firms. Few companies will have the expertise to create an AI risk management system internally.

[81] According to the Three Lines of Defence (3LoD) model, the first line, i.e. operational management, would ultimately be responsible for establishing the risk management system. However, the second line, especially the risk management team, would typically be the ones who actually create the policies, procedures, and instructions. For more information on the 3LoD model, see Institute of Internal Auditors (IIA), 'The Three Lines of Defense in Effective Risk Management and Control' (2013) <https://perma.cc/NQM2-DD7V>; IIA, 'The IIA's Three Lines Model: An Update of the Three Lines of Defense' (2020) <https://perma.cc/GAB5-DMN3>. For more information on the 3LoD model in an AI context, see Jonas Schuett, 'Three Lines of Defense against Risks from AI' (forthcoming).

[82] See the description of the implementation process in Clause 5.5 of 'ISO 31000:2018 Risk Management — Guidelines', supra, note 7.

[83] This formulation is taken from the documentation requirements of the quality management system in Art. 17(1), sentence 2, point (g). Arguably, the terms should be interpreted similarly in both cases.

[84] See Art. 16, point (j).

[85] See Art. 9(2), sentence 1.

[86] The term "AI system" is defined in Art. 3, point 1, while Art. 6 and Annex III specify which AI systems qualify as high-risk. See also Section IV.

[87] See Section V.3.

*a. Identification and analysis of known and foreseeable risks, Article 9(2), sentence 2, point (a)*

First, risks need to be identified and analysed.[89] "Risk identification" means the systematic use of available information to identify hazards,[90] whereas "hazard" can be defined as a "potential source of harm".[91] Since the AI Act does not specify how providers should identify risks, they have to rely on existing techniques and methods (e.g. risk taxonomies,[92] incident databases,[93] or scenario analysis[94]).[95] It is unclear what the AI Act means by "risk analysis". The term typically refers to both risk identification and risk estimation,[96] but this does not make sense in this context, as both steps are described separately. To avoid confusion, the legislator should arguably remove the term "analysis" from Article 9, sentence 2, point (a), or adjust point (b), as has been suggested by the French Presidency of the Council[97] and adopted by the Czech Presidency[98] (see Section V.2.b).

---

[88] 'ISO/IEC Guide 51:2014 Safety Aspects — Guidelines for Their Inclusion in Standards' <https://www.iso.org/standard/53940.html> accessed 2 November 2022.

[89] Art. 9(2), sentence 2, point (a).

[90] See Clause 3.10 and 6.1 of 'ISO/IEC Guide 51:2014 Safety Aspects — Guidelines for Their Inclusion in Standards', supra, note 88; see also Clause 3.5.1 of 'ISO Guide 73:2009 Risk Management — Vocabulary' <https://www.iso.org/standard/44651.html> accessed 2 November 2022.

[91] Clause 3.2 of 'ISO/IEC Guide 51:2014 Safety Aspects — Guidelines for Their Inclusion in Standards', supra, note 88; see also Clause 3.5.1.4 of 'ISO Guide 73:2009 Risk Management — Vocabulary', supra, note 90.

[92] E.g. Microsoft, 'Types of Harm' (2022) <https://perma.cc/FE26-NJCT>; Laura Weidinger and others, 'Ethical and Social Risks of Harm from Language Models' (arXiv, 2021) <https://arxiv.org/abs/2112.04359>; Inioluwa D Raji and others, 'The Fallacy of AI Functionality' (ACM Conference on Fairness, Accountability, and Transparency, Seoul, 2022) <https://doi.org/10.1145/3531146.3533158>.

[93] E.g. the AI Incident Database (Sean McGregor, 'Preventing Repeated Real World AI Failures by Cataloging Incidents: The AI Incident Database' [arXiv, 2020] <https://arxiv.org/abs/2011.08512>) or the OECD Global AI Incidents Tracker (OECD, 'OECD Framework for the Classification of AI Systems' [2022] 66 <https://doi.org/10.1787/cb6d9eca-en>), which is currently under development.

[94] See Luciano Floridi and Andrew Strait, 'Ethical Foresight Analysis: What it is and Why it is Needed?' (2020) 30 Minds and Machines 77 <https://doi.org/10.1007/s11023-020-09521-y>.

[95] For an overview of risk identification techniques, see Clauses B.2 and B.3 of 'IEC 31010:2019 Risk Management — Risk Assessment Techniques' <https://www.iso.org/standard/72140.html> accessed 2 November 2022.

[96] See Clause 3.10 of 'ISO/IEC Guide 51:2014 Safety Aspects — Guidelines for Their Inclusion in Standards', supra, note 88.

Risk identification and analysis should be limited to "the known and foreseeable risks associated with each high-risk AI system". However, the AI Act does not define the term "risk", nor does it say when risks are "known" or "foreseeable". I suggest using the following definitions.

"Risk" is the "combination of the probability of occurrence of harm and the severity of that harm",[99] "harm" means any adverse effect on health, safety and fundamental rights,[100] while the "probability of occurrence of harm" is "a function of the exposure to [a] hazard, the occurrence of a hazardous event, [and] the possibilities of avoiding or limiting the harm".[101] It is worth noting, however, that these definitions are not generally accepted and that there are competing concepts of risk.[102] In addition, the French Presidency of the Council has suggested a clarification,[103] which the Czech Presidency has adopted,[104] according to which the provision only refers to risks "most likely to occur to health, safety and fundamental rights in view of the intended purpose of the high-risk AI system".[105]

A risk is "known" if the harm has occurred in the past or is certain to occur in the future. To avoid circumventions, "known" refers to what an organisation could know with reasonable effort, not what they actually know. For example, a risk should be considered known if there is a relevant entry in one of the incident databases,[106] or if a public incident report has received significant media attention.

---

[97] French Presidency of the Council, supra, note 24.

[98] The Czech Presidency of the Councils, supra, note 25.

[99] Clause 3.9 of 'ISO/IEC Guide 51:2014 Safety Aspects — Guidelines for Their Inclusion in Standards', supra, note 88.

[100] According to the explanatory memorandum, risks should "be calculated taking into account the impact on rights and safety" (COM [2021] 206 final, supra, note 1, 8). See also my discussion of the purpose of Art. 9 in Section III, and the definition of "harm" in Clause 3.1 of 'ISO/IEC Guide 51:2014 Safety Aspects — Guidelines for Their Inclusion in Standards', supra, note 88.

[101] Clause 5 of 'ISO/IEC Guide 51:2014 Safety Aspects — Guidelines for Their Inclusion in Standards', supra, note 88. The terms "hazard", "hazardous event", and "hazardous situation" are defined in Clauses 3.2–3.4.

[102] E.g. the term "risk" can also be defined as an "effect of uncertainty on objectives" (Clause 3.1 of 'ISO 31000:2018 Risk Management — Guidelines', supra, note 7). For more information on the different concepts of risk, see Mahler, supra, note 19, 256–260; see also Margot E Kaminski, 'Regulating the Risks of AI' 103 Boston University Law Review (forthcoming) <http://dx.doi.org/10.2139/ssrn.4195066>.

[103] French Presidency of the Council, supra, note 24.

[104] The Czech Presidency of the Council, supra, note 25.

[105] The reference to health, safety and fundamental rights seems to clarify the purpose of the norm (see Section IV), while the reference to the intended purpose seems to be a consequence of deleting point (b) (see Section V.2.b).

A risk is "foreseeable" if it has not yet occurred but can already be identified. The question of how much effort organisations need to put into identifying new risks involves a difficult trade-off. On the one hand, providers need legal certainty. In particular, they need to know when they are allowed to stop looking for new risks. On the other hand, the AI Act should prevent situations where providers cause significant harm, but are able to exculpate themselves by arguing the risk was not foreseeable. If this were possible, the AI Act would fail to protect health, safety and fundamental rights. A possible way to resolve this trade-off is the following rule of thumb: the higher the potential impact of the risk, the more effort an organisation needs to put into foreseeing it. For example, it should be extremely difficult for a provider to credibly assure that a catastrophic risk was unforeseeable.[107]

*b. Estimation and evaluation of risks that may emerge from intended uses or foreseeable misuses, or risks that have been identified during post-market monitoring, Article 9(2), sentence 2, points (b), (c)*

Next, risks need to be estimated and evaluated.[108] "Risk estimation" means the estimation of the probability of occurrence of harm and the severity of that harm.[109] Since the AI Act does not specify how to estimate risks, providers have to rely on existing techniques (e.g. Bayesian networks and influence diagrams).[110] "Risk evaluation" means the determination of whether a risk is acceptable.[111] I discuss this question in more detail below (see Section V.3).

Risk estimation and evaluation should only cover risks "that may emerge when the high-risk AI system is used in accordance with its intended purpose and under conditions of reasonably foreseeable misuse".[112] The terms "intended purpose" and "reasonable foreseeable misuse" are both defined in the AI Act.[113] If the system is not used as intended or misused in an unforeseeable way, the risks do not have to be included. This ensures that the provider is only responsible for risks they can control, which increases legal certainty. To prepare this step, providers should identify potential users, intended uses, and reasonably foreseeable misuses at the beginning of the risk management process.[114]

---

[106] E.g. the AI Incident Database (Sean McGregor, 'Preventing Repeated Real World AI Failures by Cataloging Incidents: The AI Incident Database' [arXiv, 2020] <https://arxiv.org/abs/2011.08512>) or the OECD Global AI Incidents Tracker (OECD, 'OECD Framework for the Classification of AI Systems' [2022] 66 <https://doi.org/10.1787/cb6d9eca-en>), which is currently under development.

[107] For more information on addressing catastrophic risks through AI risk management measures, see Anthony M Barrett and others, 'Actionable Guidance for High-Consequence AI Risk Management: Towards Standards Addressing AI Catastrophic Risks' (arXiv, 2022) 9 <https://arxiv.org/abs/2206.08966>.

[108] Art. 9(2), sentence 2, point (b).

[109] See Clauses 3.9 and 3.10 of 'ISO/IEC Guide 51:2014 Safety Aspects — Guidelines for Their Inclusion in Standards', supra, note 88; see also the other definitions in Clause 3.

[110] For an overview of risk estimation techniques, see Clauses B.5 and B.8 of 'IEC 31010:2019 Risk Management — Risk Assessment Techniques', supra, note 95. See also Microsoft, 'Foundations of Assessing Harm' (2022) <https://perma.cc/7H6P-UDM7>.

[111] See Clause 3.12 of 'IEC 31010:2019 Risk Management — Risk Assessment Techniques', supra, note 95.

Providers of high-risk AI systems also need to evaluate risks that they have identified through their post-market monitoring system.[115] This provision ensures that providers also manage risks from unintended uses or unforeseeable misuses if they have data that such practices exist. While this expands the circle of relevant risks, it does not threaten legal certainty.

Note that the French Presidency of the Council,[116] followed by the Czech Presidency,[117] has proposed to delete Article 9(2), sentence 2, point (b) and to add a sentence 3 instead: "The risks referred to in [paragraph 2] shall concern only those which may be reasonably mitigated or eliminated through the development or design of the high-risk AI system, or the provision of adequate technical information." These changes would limit the types of risks that providers of AI systems are responsible for compared to the original proposal by the EC.

*c. Adoption of risk management measures, Article 9(2), sentence 2, point (d)*

Finally, suitable risk management measures need to be adopted.[118] "Risk management measures" (also known as "risk response" or "risk treatment") are actions that are taken to reduce the identified and evaluated risks. Paragraphs 3 and 4 contain more details about specific measures (see Section V.3).

Although the three steps are presented in a sequential way, they are meant to be "iterative".[119] As alluded to in Section II, the risk management process needs to be repeated until all risks have been reduced to an acceptable level. After the first two steps, providers need to decide if the risk is already acceptable. If this is the case, they can document their decision and complete the process. Otherwise, they need to move on to the third step. After they have adopted suitable risk management measures, they need to reassess the risk and decide if the residual risk is acceptable. If it is not, they have to take additional risk management measures. If it turns out that it is not possible to reduce residual risks to an acceptable level, the development and deployment process must be stopped.[120] Although the AI Act does not reference it, the iterative process described in paragraph 2 is very similar to the one described in ISO/IEC Guide 51.[121] It is illustrated in Figure 1.

---

[112] Art. 9(2), sentence 2, point (b).
[113] Art. 3, points 12 and 13.
[114] Similar to Clause 6.1 of 'ISO/IEC Guide 51:2014 Safety Aspects — Guidelines for Their Inclusion in Standards', supra, note 88.
[115] Art. 9(2), sentence 2, point (c). The post-market monitoring system is described in Art. 61.
[116] French Presidency of the Council, supra, note 24.
[117] Czech Presidency of the Council, supra, note 25.
[118] Art. 9(2), sentence 2, point (d).
[119] Art. 9(2), sentence 1.

The risk management process needs to "run throughout the entire lifecycle of a high-risk AI system".[122] The original EC proposal does not define "lifecycle of an AI system", but the French Presidency of the Council has suggested a new definition,[123] which the Czech Presidency has adopted.[124] (According to the Czech Presidency, the risk management process also needs to be "planned" throughout the entire lifecycle.[125]) In practice, providers will need to know how often and when in the lifecycle they must complete the risk management process. In the absence of an explicit requirement, providers have to rely on considerations of expediency. They should perform a first iteration early on in the development process and, based on the findings of that iteration, decide how to proceed. For example, if they only identify a handful of low-probability, low-impact risks, they may decide to run fewer and less thorough iterations later in the life cycle. However, two iterations, one during the development stage and one before deployment,[126] seems to be the bare minimum.

---

[120] Art. 9 does not say this explicitly, but it seems to be a logical consequence of the process.

[121] See Clause 6.1 of 'ISO/IEC Guide 51:2014 Safety Aspects — Guidelines for Their Inclusion in Standards', supra, note 88.

[122] Art. 9(2), sentence 1.

[123] French Presidency of the Council, supra, note 24.

[124] The Czech Presidency of the Council, supra, note 25 defines "lifecycle of an AI system" as "the duration of an AI system, from design through retirement. Without prejudice to the powers of the market surveillance authorities, such retirement may happen at any point in time during the post-market monitoring phase upon the decision of the provider and implies that the system may not be used further. An AI system lifecycle is also ended by a substantial modification to the AI system made by the provider or any other natural or legal person, in which case the substantially modified AI system shall be considered as a new AI system." See also the AI system lifecycle model from OECD, 'Scoping the OECD AI Principles: Deliberations of the Expert Group on Artificial Intelligence at the OECD (AIGO)' (2019) 13 <https://doi.org/10.1787/d62f618a-en>, which distinguishes between four stages: (1) design, data and modelling, (2) verification and validation, (3) deployment, and (4) operation and monitoring. See also the modified version from NIST, supra, note 13, 5.

[125] Czech Presidency of the Council, supra, note 25.

[126] This is similar to the testing requirements set out in Art. 9(7), according to which testing "shall be performed, as appropriate, at any point in time throughout the development process, and, in any event, prior to the placing on the market or the putting into service."

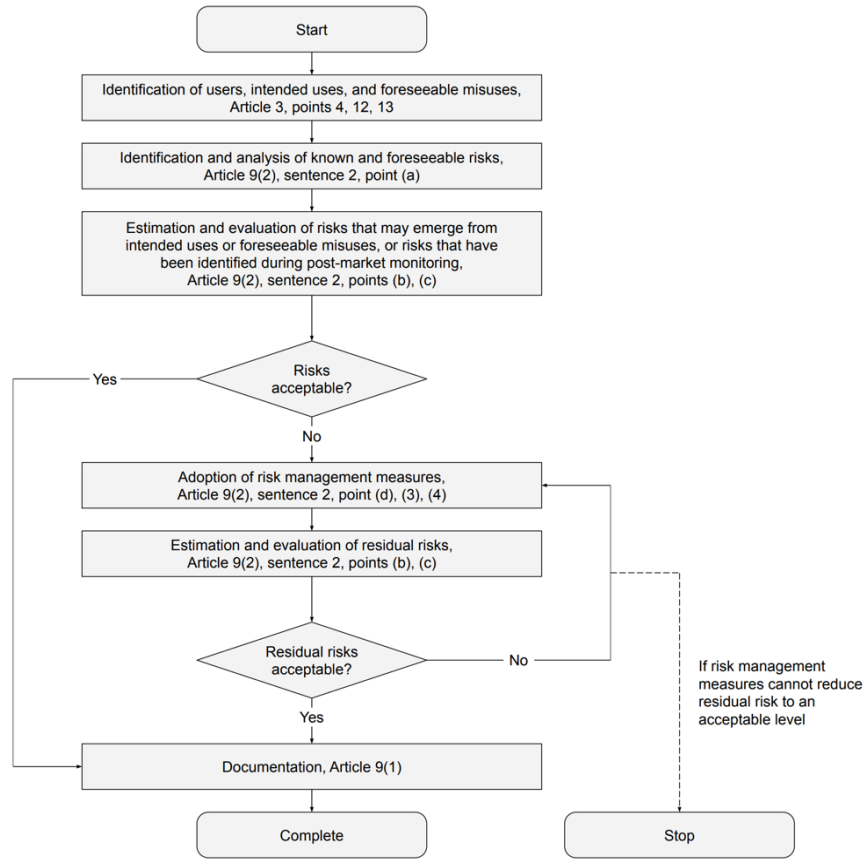

*Figure 1:* Overview of the risk management process described in Article 9(2) based on the iterative process of risk assessment and risk reduction described in ISO/IEC Guide 51.[127]

*3. Risk management measures, Article 9(3), (4)*

Paragraphs 3 and 4 contain more details about the risk management measures referred to in paragraph 2, sentence 2, point (d). According to paragraph 3, the risk management measures "shall give due consideration to the effects and possible interactions resulting from the combined application of the requirements set out in […] Chapter 2".[128] Besides that, they "shall take into account the

---

[127] Clause 6.1 of 'ISO/IEC Guide 51:2014 Safety Aspects — Guidelines for Their Inclusion in Standards', supra, note 88.

[128] Note that the French Presidency of the Council, supra, note 24, followed by the Czech Presidency, supra, note 25, has proposed to add the following half sentence: "with a view to

generally acknowledged state of the art, including as reflected in relevant harmonised standards or common specifications".[129] It is worth noting that there are not yet any harmonised standards[130] or common specifications[131] on AI risk management. It is probably also too early for a "generally acknowledged state of the art", but emerging AI risk management standards[132] and ERM frameworks[133] could serve as a starting point.

Paragraph 4 contains three subparagraphs. The first specifies the purpose of adopting risk management measures, the second lists specific measures, and the third is about the socio-technical context.

The purpose of adopting risk management measures is to reduce risks "such that any residual risk […] is judged acceptable". A "residual risk" is any "risk remaining after risk reduction measures have been implemented".[134] "Acceptable risk" (or "tolerable risk") can be defined as the "level of risk that is accepted in a given context based on the current values of society".[135] To make this definition more concrete, it could be interpreted in light of the purpose of the norm (see Section III). The "current values of society" would then entail a high level of protection of public interests, especially health, safety and fundamental rights. In addition to that, providers may want to consider their own risk appetite,[136] as required by most ERM systems. It is worth noting, however, that defining normative thresholds is still an open problem in AI ethics,[137] both for individual characteristics (e.g. how fair is fair enough?) and trade-offs between different characteristics (e.g. increasing fairness might reduce privacy).[138] Until harmonised standards provide further guidance, providers will have to use

---

minimising risks more effectively while achieving an appropriate balance in implementing the measures to fulfil those requirements".

[129] As mentioned in Section II, the French Presidency of the Council, supra, note 24, followed by the Czech Presidency of the Council, supra, note 25, has suggested deleting this sentence. Note that this would not undermine the importance of harmonised standards and common specifications due to the presumption of conformity in Art. 40.

[130] The term "harmonised standard" is defined in Art. 3, point 27.

[131] The term "common specifications" is defined in Art. 3, point 28.

[132] E.g. NIST, supra, note 13; 'ISO/IEC FDIS 23894 Information Technology — Artificial Intelligence — Guidance on Risk Management', supra, note 14.

[133] E.g. 'ISO 31000:2018 Risk Management — Guidelines', supra, note 7; COSO, 'Enterprise Risk Management—Integrating with Strategy and Performance', supra, note 15.

[134] Clause 3.8 of 'ISO/IEC Guide 51:2014 Safety Aspects — Guidelines for Their Inclusion in Standards', supra, note 88.

[135] Clause 3.15 of ibid.

[136] The term "risk appetite" can be defined as the "amount and type of risk that an organization is willing to pursue or retain" (Clause 3.7.1.2 of 'ISO Guide 73:2009 Risk Management — Vocabulary', supra, note 90).

[137] See Brent Mittelstadt, 'Principles Alone Cannot Guarantee Ethical AI' (2019) 1 Nature Machine Intelligence 501 <https://doi.org/10.1038/s42256-019-0114-4>.

[138] See Bryce Goodman, 'Hard Choices and Hard Limits in Artificial Intelligence' (2021) <https://doi.org/10.1145/3461702.3462539>.

their own definitions or rely on popular definitions from others. Paragraph 4, subparagraph 1 further states that "each hazard as well as the overall residual risk" must be judged acceptable. In other words, providers must consider risks both individually and collectively, but only if the system "is used in accordance with its intended purpose or under conditions of reasonably foreseeable misuse".[139] Finally, "those residual risks [that are judged acceptable] shall be communicated to the user".[140]

Providers of high-risk AI systems must adopt three types of risk management measures. These measures resemble the "three-step-method" in ISO/IEC Guide 51.[141] First, providers must design and develop the system in a way that eliminates or reduces risks as much as possible.[142] For example, to reduce the risk that a language model outputs toxic language,[143] providers could fine-tune the model.[144] Second, if risks cannot be eliminated, providers must implement adequate mitigations and control measures, where appropriate.[145] If fine-tuning the language model is not enough, the provider could use safety filters[146] or other approaches to content-detection.[147] Third, they must provide adequate information and, where appropriate, training to users.[148] Figure 2 gives an overview of the three types of measures and illustrates how they collectively reduce risk.

---

[139] The terms "intended purpose" and "reasonably foreseeable misuse" are defined in Art. 3, points 12, 13. Note that the French Presidency of the Council, supra, note 24, and the Czech Presidency, supra, note 25, have suggested deleting this requirement.

[140] This requirement should be read in conjunction with Art. 9(4), subparagraph 2, point (c) and Art. 13. Note that the French Presidency of the Council, supra, note 24, followed by the Czech Presidency of the Council, supra, note 25, has suggested deleting this requirement.

[141] See Clauses 6.3.4 and 6.3.5 of 'ISO/IEC Guide 51:2014 Safety Aspects — Guidelines for Their Inclusion in Standards', supra, note 88. The three steps are "(1) inherently safe design; (2) guards and protective devices; (3) information for end users". It is worth noting, however, that the AI Act only specifies risk reduction measures for the design phase; it does not specify any measures for the use phase.

[142] Art. 9(4), subparagraph 2, point (a). Note that the French Presidency of the Council, supra, note 24, and the Czech Presidency of the Council, supra, note 25, have suggested a clarification, according to which the provision only refers to "identified and evaluated" risks.

[143] For more information on this type of risk, see Laura Weidinger and others, 'Ethical and Social Risks of Harm from Language Models' (arXiv, 2021) 15–16 <https://arxiv.org/abs/2112.04359>.

[144] E.g. Irene Solaiman and Christy Dennison, 'Process for Adapting Language Models to Society (PALMS) with Values-Targeted Datasets' (35th Annual Conference on Advances in Neural Information Processing Systems, Virtual, 2021) <https://perma.cc/DR9G-X69X>.

[145] Art. 9(4), subparagraph 2, point (b).

[146] See e.g. Javier Rando and others, 'Red-Teaming the Stable Diffusion Safety Filter' (arXiv 2022) <https://arxiv.org/abs/2210.04610>.

[147] See e.g. Todor Markov and others, 'A Holistic Approach to Undesired Content Detection in the Real World' (arXiv 2022) <https://arxiv.org/abs/2208.03274>.

[148] Art. 9(4), subparagraph 2, point (c); see also Art. 13.

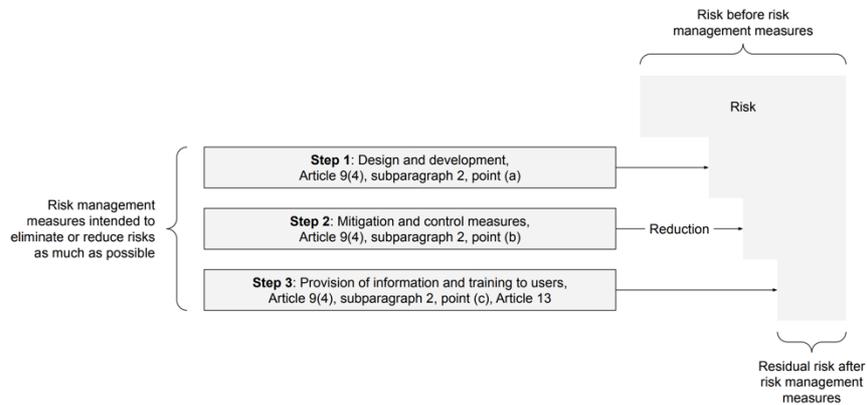

*Figure 2:* Overview of risk management measures described in Article 9(4), subsection 2 inspired by ISO/IEC Guide 51.[149]

Finally, when adopting the above-mentioned risk management measures to reduce risks related to the use of the system, providers must give "due consideration […] to the technical knowledge, experience, education, training to be expected by the user and the environment in which the system is intended to be used." The provision acknowledges that AI systems are always embedded in their socio-technical context.

### 4. Testing procedures, Article 9(5)-(7)

The second component of the risk management system are testing procedures. Pursuant to paragraph 5, sentence 1, "high-risk AI systems shall be tested". "Testing" can be defined as a "set of activities conducted to facilitate discovery and evaluation of properties of the test items".[150] This typically involves the use of metrics and probabilistic thresholds.[151] Below, I discuss the "why", "when", "how", and "who" of testing.

Pursuant to paragraph 5, testing has three purposes. First, it is aimed at "identifying the most appropriate risk management measures".[152] Let us revisit our example of a language model that outputs toxic language. While providers could take many different measures to reduce that risk, testing (e.g. using toxicity classifiers[153]) can give them a better understanding of the risk and thereby help them adopt more appropriate measures. (However, the Czech Presidency of the Council has suggested dropping this first part of the provision.[154]) Second, testing shall "ensure that high-risk AI systems perform consistently for their intended purpose".[155] AI systems often perform worse when the environment in which they are actually used differs from their training environment. This problem is known as "distributional shift".[156] Testing can help providers detect when it is particularly likely that the system will perform poorly in the environment it is intended for (so-called "out-of-distribution detection"). Third, testing shall ensure that high-risk AI systems "are in compliance with the requirements set out in [Chapter 2]".[157] Some of these provisions require the system to have certain properties like being "sufficiently transparent"[158] or having "an appropriate level of accuracy, robustness and cybersecurity".[159] Testing can evaluate how well the system performs on these dimensions relative to certain benchmarks, helping providers interpret whether the current level is in fact "sufficient" or "appropriate".[160]

---

[149] Clause 6.3.4 of 'ISO/IEC Guide 51:2014 Safety Aspects — Guidelines for Their Inclusion in Standards', supra, note 88.

[150] Clause 3.131 of 'ISO/IEC/IEEE 29119-1:2022 Software and Systems Engineering — Software Testing — Part 1: General Concepts' <https://www.iso.org/standard/81291.html> accessed 2 November 2022.

[151] See Art. 9(7), sentence 2.

[152] Art. 9(5), sentence 1. See also Art. 9(2), sentence 2, point (d), (3), (4), and Sections V.2, V.3.

Paragraph 6 only refers to "AI systems", not "high-risk AI systems", but this seems to be the result of a mistake in the drafting of the text. The provision states that testing procedures "shall be suitable to achieve the intended purpose" and not "go beyond what is necessary to achieve that purpose". This is essentially a restatement of the principle of proportionality. Besides that, the paragraph does not seem to have a discrete regulatory content. Presumably in light of this, the French[161] and Czech Presidency of the Council[162] have proposed to substitute the provision with a reference to a new Article 54a that lays out rules on testing in real world conditions.

---

[153] E.g. 'Perspective API' (*GitHub*) <https://github.com/conversationai/perspectiveapi> accessed 1 November 2022.

[154] Czech Presidency of the Council, supra, note 25.

[155] Art. 9(5), sentence 2. The term "intended purpose" is defined in Art. 3, point 12.

[156] For more information on the problem of distributional (or dataset) shift, see Joaquin Quiñonero-Candela and others (eds), *Dataset Shift in Machine Learning* (MIT Press 2022). See also Amodei and others, supra, note 8, 16–20.

[157] Art. 9(5), sentence 2. In addition to Art. 8 and 9, Chapter 2 contains requirements on data and data governance (Art. 10), technical documentation (Art. 11), record-keeping (Art. 12), transparency and provision of information to users (Art. 13), human oversight (Art. 14), and accuracy, robustness and cybersecurity (Art. 15).

[158] Art. 13(1), sentence 1.

[159] Art. 15(1).

[160] Chapter 2 contains both technical requirements for high-risk AI systems (e.g. regarding their accuracy) and governance requirements for the providers of such systems (e.g. regarding record-keeping). Although paragraph 5 refers to both types of requirements, it only makes sense for technical requirements. For example, there do not seem to be any metrics or probabilistic thresholds for documentation (Art. 11) or record-keeping (Art. 12).

[161] French Presidency of the Council, supra, note 24.

[162] Czech Presidency of the Council, supra, note 25.

Paragraph 7, sentence 1 specifies *when* providers must test their high-risk AI systems, namely "as appropriate, at any point in time throughout the development process, and, in any event, prior to the placing on the market or the putting into service." Note that this is different from the risk management process (see Section V.2). While the risk management process needs to "run through the entire lifecycle",[163] testing only needs to be performed "throughout the development process". Although the formulation "as appropriate" indicates that providers have discretion when and how often to test their systems, testing must be performed "prior to the placing on the market or the putting into service".[164]

Paragraph 7, sentence 2 specifies *how* providers must test their high-risk AI systems, namely "against preliminarily defined metrics and probabilistic thresholds that are appropriate to the intended purpose of the high-risk AI system". "Metric" includes assessment criteria, benchmarks, and key performance indicators (KPIs). "Probabilistic thresholds" are a special kind of metric that evaluate a property on a probabilistic scale with one or more predefined thresholds. It is not possible to make any general statements as to which metric or probabilistic threshold to use, mainly because their appropriateness is very context-specific and because there are not yet any best practices. Providers will therefore have to operate under uncertainty and under the assumption that metrics they have used in the past might not be appropriate in the future. Presumably, this is the reason why the norm speaks of "preliminary defined metrics".

The norm does not specify *who* must perform the testing. As discussed in Section IV, it applies to providers of high-risk AI systems. But do providers need to perform the testing themselves, or can they outsource it? I expect that many providers want to outsource the testing or parts thereof (e.g. the final testing before placing the system on the market). In my view, this seems to be unproblematic, as long as the provider remains responsible for meeting the requirements.[165]

*5. Special rules for children and credit institutions, Article 9(8), (9)*

Paragraph 8 contains special rules for children. (The French Presidency of the Council has specified this as "persons under the age of 18",[166] and the Czech Presidency has adopted that suggestion.[167]) When implementing the risk management system, "specific consideration shall be given to whether the high-risk AI system is likely to be accessed by or have an impact on children". Children take a special role in the AI Act because they are particularly vulnerable and have specific rights.[168] Providers of high-risk AI systems must therefore take special measures to protect them.

---

[163] Art. 9(2), sentence 1.

[164] The terms "placing on the market" and "putting into service" are defined in Art. 3, points 9 and 11.

[165] If the outsourcing company does not perform the testing in accordance with Art. 9(5)-(7), the provider would still be subject to administrative and civil enforcement measures (see Section VI). The provider could only claim recourse from the outsourcing company.

[166] French Presidency of the Council, supra, note 24.

[167] Czech Presidency of the Council, supra, note 25.

Paragraph 9 contains a collusion rule for credit institutions. Since credit institutions are already required to implement a risk management system,[169] one might ask how the AI-specific requirements relate to the credit institution-specific ones. Paragraph 9 clarifies that the AI-specific requirements "shall be part" of the credit institution-specific ones. In other words, Article 9 complements existing risk management systems, it does not replace them. In light of this, the Czech Presidency of the Council has suggested extending paragraph 9 to any provider of high-risk AI systems that is already required to implement a risk management system.[170]

But what happens if providers of high-risk AI systems do not comply with these requirements? The next section gives an overview of possible enforcement mechanisms.

## VI. Enforcement

In this section, I describe ways in which Article 9 can be enforced. This might include administrative, civil, and criminal enforcement measures.

Providers of high-risk AI systems that do not comply with Article 9 can be subject to administrative fines of up to € 20 million or, if the offender is a company, up to 4% of its total worldwide annual turnover for the preceding financial year, whichever is higher.[171] The French Presidency of the Council[172] followed by the Czech Presidency[173] proposed to limit this fine in case of a small and medium-sized enterprise (SME) to 2% of its total worldwide annual turnover for the preceding financial year. The AI Act only contains high-level guidelines on penalties (e.g. how to decide on the amount of administrative fines[174]), the details will be specified by each member state.[175] In practice, I expect administrative fines to be significantly lower than the upper bound, similar to the GDPR.[176] Before imposing penalties and administrative fines, national competent authorities[177] will usually request providers of high-risk AI systems to demonstrate conformity with the requirements set out in Article 9.[178] Supplying incorrect, incomplete or misleading information can entail further administrative fines.[179]

---

[168] See Recital 28. For more information on the potential impact of AI systems on children, see Vasiliki Charisi and others, *Artificial Intelligence and the Rights of the Child: Towards an Integrated Agenda for Research and Policy* (Publications Office of the European Union 2022) <http://dx.doi.org/10.2760/012329>.
[169] See Art. 74 of the Directive 2013/36/EU.
[170] Czech Presidency of the Council, supra, note 25.
[171] See Art. 71(4).
[172] French Presidency of the Council, supra, note 24; see also Bertuzzi, supra, note 44.
[173] Czech Presidency of the Council, supra, note 25.
[174] See Art. 71(6).
[175] See Art. 71(1).

Providers of high-risk AI systems might also be subject to civil liability. First, the provider might be held contractually liable. If a contracting party of the provider is harmed, then this party might claim compensation from the provider. This will often depend on the question if complying with Article 9 is a contractual accessory obligation. Second, there might be a tort law liability. If a high-risk AI system harms a person, that person may claim compensation from the provider of that system. In some member states, this will largely depend on the question whether Article 9 protects individuals (see Section III).[180] Third, there might be an internal liability. If a company has been fined, it might claim recourse from the responsible manager.[181] This mainly depends on the question if not implementing a risk management system can be seen as a breach of duty of care.

Finally, Article 9 is not directly enforceable by means of criminal law. Although the AI Act does not mention any criminal enforcement measures, violating Article 9 might still be an element of a criminal offence in some member states. For example, a failure to implement a risk management system might constitute negligent behaviour.[182]

## VII. Conclusion

This article has analysed Article 9, the key risk management provision in the AI Act. Section II gave an overview of the regulatory concept behind the norm. I argued that Article 9 shall ensure that providers of high-risk AI systems identify risks that remain even if they comply with the other requirements set out in Chapter 2, and take additional measures to reduce them. Section III determined the purpose of Article 9. It seems uncontroversial that the norm is intended to improve the functioning of the internal market and protect the public interest. But I also raised the question whether the norm also protects certain individuals. Section IV determined the norm's scope of application. Materially and personally, Article 9 applies to providers of high-risk AI systems. Section V offered a comprehensive interpretation of the specific risk management requirements. Paragraph 1 contains the central requirement, according to which providers of high-risk AI systems must implement a risk management system, while paragraphs 2 to 7 specify the details of that system. The iterative risk management process is illustrated in Figure 1, while Figure 2 shows how different risk management measures can collectively reduce risk. Paragraphs 8 and 9 contain special rules for children and credit institutions. Section VI described ways in which these requirements can be enforced, in particular via penalties and administrative fines as well as civil liability.

---

[176] The upper bound of administrative fines in the GDPR is similarly high, see Art. 84 of the GDPR. However, findings of a recent study suggest that, in practice, the majority of fines only range from a few hundreds to a few hundred thousand euros (Jukka Ruohonen and Kalle Hjerppe, 'The GDPR Enforcement Fines at Glance' [2022] 106 Information Systems 101876 <https://doi.org/10.1016/j.is.2021.101876>).

[177] The term "national competent authority" is defined in Art. 3, point 43.

[178] See Art. 16(j) and Art. 23, sentence 1.

[179] See Art. 71(5).

[180] See e.g. Section 823(2) of the German Civil Code.

[181] See e.g. Section 93(2), sentence 1 of the German Stock Corporation Act, or Section 43(2) of the German Limited Liability Companies Act.

[182] See Mihailis E Diamantis, 'The Extended Corporate Mind: When Corporations Use AI to Break the Law' (2020) 97 North Carolina Law Review 893 <https://perma.cc/RP8T-BSZL>.

Based on my analysis in Section V, I suggest three amendments to Article 9 (or specifications in harmonised standards). First, I suggest adding a passage on the organisational dimension of risk management, similar to the Govern function in the NIST AI Risk Management Framework,[183] which is compatible with existing best practices like the Three Lines of Defence (3LoD) model.[184] Second, I suggest adding a requirement to evaluate the effectiveness of the risk management system. The most obvious way to do that would be through an internal audit function. Third, I suggest clarifying that the risk management system is intended to reduce individual, collective, and societal risks,[185] not just risks to the provider of high-risk AI systems.

The article makes three main contributions. First, by offering a comprehensive interpretation of Article 9, it helps providers of high-risk AI systems to comply with the risk management requirements set out in the AI Act. Although it will take several years until compliance is mandatory, they may want to know as early as possible what awaits them. Second, the article has suggested ways in which Article 9 can be amended. And third, it informs future efforts to develop harmonised standards on AI risk management in the EU.

Although my analysis focuses on the EU, I expect it to be relevant for policy makers worldwide. In particular, it might inform regulatory efforts in the US[186]

---

[183] NIST, supra, note 13, 18–19.

[184] For more information on the 3LoD model, see IIA, supra, note 81. For more information on the 3LoD model in an AI context, see Jonas Schuett, supra, note 81.

[185] See Nathalie A Smuha, 'Beyond the Individual: Governing AI's Societal Harm' (2021) 10 Internet Policy Review <https://doi.org/10.14763/2021.3.1574>.

[186] The White House, 'Guidance for Regulation of Artificial Intelligence Applications' (2020) 4 <https://perma.cc/U2V3-LGV6> explicitly mentions risk assessment and management in a regulatory context. It also seems plausible that the NIST AI Risk Management Framework (NIST, supra, note 13) will be translated into law, similar to the NIST

and UK,[187] especially since risk management as a governance tool is not inherently tied to EU law and there is value in compatible regulatory regimes.

## Acknowledgements

I am grateful for valuable comments and feedback from Leonie Koessler, Markus Anderljung, Christina Barta, Christoph Winter, Robert Trager, Noemi Dreksler, Eoghan Stafford, Jakob Mökander, Elliot Jones, Andre Barbe, Risto Uuk, Alexandra Belias, Haydn Belfied, Anthony Barrett, James Ginns, and Emma Bluemke. I also thank the participants of a seminar hosted by the Centre for the Governance of AI in July 2022. All remaining errors are my own.

---

Cybersecurity Framework (NIST, 'Framework for Improving Critical Infrastructure Cybersecurity: Version 1.1' <https://perma.cc/JC5V-6YNS>).

[187] The UK National AI Strategy promised a white paper on AI regulation, which the Office for AI intends to publish in 2022 (HM Government, 'National AI Strategy' [2021] 53 <https://perma.cc/RYN4-EEBR>). As a first step towards this white paper, the Department for Digital, Culture, Media & Sport (DCMS), Department for Business, Energy & Industrial Strategy (BEIS), and Office for AI published a policy paper (DCMS, BEIS and Office for AI, 'Establishing a Pro-Innovation Approach to Regulating AI - An Overview of the UK's Emerging Approach' [2022] <https://perma.cc/VG25-XAEZ>). Although both documents do not explicitly mention risk management, I expect the final regulation to contain provisions on risk management.